\documentclass[12pt]{article}
\newcommand {\be}{\begin{equation}}
\newcommand {\ee}{\end{equation}}
\usepackage{graphicx}
\begin{document}

\title{Geometry and Dynamics of Quantum State Diffusion}
 \author{Nikola Buri\' c  \thanks{e-mail: buric@phy.bg.ac.yu}
\\
Institute of Physics,
 P.O. Box 57, 11001 Belgrade, Serbia.}

 \maketitle

\begin{abstract}
 Riemannian metric on real 2n-dimensional space  associated with the equation
 governing  complex diffusion of pure states of an open quantum system is introduced and studied.
Examples of a qubit under the influence of dephasing and thermal environments are used
 to show that  the curvature of the diffusion metric is a good indicator of
the properties of the environment dominated evolution and its
stability.

\end{abstract}


\newpage

\section{Introduction}

The states of an open quantum system are commonly described by a
density matrix $\hat \rho$. In many cases, the evolution of $\hat
\rho(t)$ is governed by a master equation of the Linblad form
\cite{Linblad},\cite{Brauer} for the
 density matrix $\hat\rho(t)$
\begin{equation}
{d\hat\rho(t)\over dt}=-i[\hat H,\hat\rho]+{1\over 2} \sum_l [\hat
L_l\hat\rho,\hat L_l^{\dag}]+[\hat L_l,\hat\rho\hat L_l^{\dag}],
\end{equation}
where the Linblad operators $\hat L_l$ describe the influence of
the environment.
 The equation (1) represents the general form of an evolution equation for a quantum system which satisfies Markov property.

 However, this theoretical approach to the dynamics of open quantum systems
 is not unique.  In real experiments it is often useful to understand and model the dynamics of
   pure quantum states \cite {Carmicele},\cite{Nature},\cite{ja1}.  Indeed, the evolution
   of an open system can be described directly in terms of the dynamics of the system's pure
    state. The corresponding evolution equation is a stochastic modification of the unitary
 Schroedinger equation. In fact, the density matrix $\hat \rho$ can be written,
in different but equivalent ways, as a convex combination of pure
states. Each of
 these results in a stochastic differential equation for $|\psi(t)>$ in the Hilbert space ${\cal H}$.
 Such stochastic Schroedinger equations (SSE) are called stochastic unravelling
 \cite{QSD},\cite{Bel},\cite{Brauer} of the
Linblad master equation for the reduced density matrix
$\hat\rho(t)$.  There are many different forms of nonlinear and
linear
  SSE that have been used in the context of open systems
   \cite{open-gen3},\cite{Carmicele}, \cite{Brauer},\cite{QSD}, \cite{Bel}
  or suggested as fundamental modifications of the Schroedinger equation \cite{Pearle}
  \cite{Girardi},\cite{Gisin},\cite{QSD}
\cite{Hugh},\cite{Adler}.
   They are all consistent with the requirement
 that the solutions of (1) and  of SSE satisfy
\begin {equation}
\hat\rho(t)=E[|\psi(t)><\psi(t)|].
\end {equation}
where $E[|\psi(t)><\psi(t)|]$ is the expectation with respect to
the distribution of the stochastic process $|\psi(t)>$.
 The advantages of the description in terms of the pure states and SSE over the description by $\hat\rho$ are twofold.
 On the practical side, the computations are much more practical, as soon as the size of the Hilbert space
  is moderate or large \cite{numerical}. On the theoretical side,  the stochastic evolution of pure states provides
   valuable insides which can not be inferred from the density
    matrix approach \cite{adv1},\cite{adv2},\cite{QSD},\cite{Brauer},\cite{ja1},\cite{ja2}.

There are two main approaches to the unravelling of the Linblad
master equation: the method of quantum state diffusion \cite{QSD}
and the relative state method \cite{Carmicele}, \cite{Brauer},
with specific advantages associated with each of the methods. The
relative state method is usually used do describe the situations
when the measurement is the dominant interaction with the
environment. The method offers particular flexibility in that the
master equation can be unravelled into different stochastic
equations conditioned on the results of measurement. On the other
hand the correspondence between the QSD equations and the Linblad
master equations is unique, and is not related to a particular
measurement scheme, or the form of the Markov environment. The
resulting SSE is always of the form of a diffusion process on the
Hilbert space of pure states, which is its main property to be
explored in this paper.

 We shall concentrate on the unique unravelling of the master equation
  given by the quantum state diffusion equation,
   and explore the fact that it represents a diffusion process.
 QSD equation is the unique  unravelling of (1) which preserves the norm of
    the state vector and  has the
same invariance as (1) under the unitary transformations of the
environment
 operators $\{\hat L_l\},  \cite{QSD}$.
 The equation is given by the following formula:
 \begin{eqnarray}
|d\psi>&=&-i\hat H |\psi>dt\nonumber\\
&+&\left [\sum_l 2<\hat L_l^{\dag
}> \hat L_l-\hat L_l^{\dag}\hat L_l
-<\hat L_l^{\dag}> <\hat L_l>\right ]|\psi(t)>dt\nonumber\\
&+& \sum_l (\hat L_l-<L_l>)|\psi(t)> dW_l
\end{eqnarray}
where $<>$ denotes the quantum expectation in the state
$|\psi(t)>$ and $dW_l$ are independent increments (indexed by $l$)
of complex Wiener c-number processes $W_l(t)$.

The equation (3) represent a diffusion process on a complex vector
space. We shall utilize  the diffusion matrix of this process to
define a Riemannian metric on the corresponding real space. We
shall then  study the properties of this diffusion metric as a field fixed by the environment and
in relation to the stochastic evolution of the state vector, for
different types of the environment.  It will be shown, using examples of the dephasing and thermal
 environments and the measurement of an observable,  that  the curvature of the diffusion metric
is a good indicator of the properties of the environment dominated evolution and its stability.
We shall see that the curvature maxima of the diffusion metric coincide with the states that are
 preferred by the particular type of the environment. Furthermore, if the maxima are sharp and positive
 the stochastic dynamics governed by the environment and  a Hamiltonian perturbation that does not commute
 with $\hat L_l$, is likely to be attracted to the state with the maximal (positive) curvature.
  On the other hand the states that correspond to the negative values of the curvature are unstable.
  Our analyzes of the QSD equation, and the results, are strictly related to the
  fact that the equation represent a norm-preserving diffusion process, and in
  this sense are applicable to the stochastic modifications of the Schroedinger equation
  that describe a norm-preserving diffusion on the Hilbert space of pure states, like the QSD equation
  and, for example, the equations
   of the spontaneous collapse models \cite{Adler}.

The structure of the paper is as follows.  We shall first discuss,
in the next section, a way to relate a Riemannian metric on a real
space $R^{2n}$ to a complex diffusion process on $C^{n}$.
Then, in section 3, we shall apply this procedure to define the
Riemannian metric associated with  QSD, and than  the properties
 of this metric for various types of environments will be studied.
  Finally, in section 4, we shall summarize and discuss our results.

\section{Riemannian metric of a complex diffusion}

Using the following notation
\begin{eqnarray}
f(|\psi>)&=&-i\hat H |\psi>\\
&+&\left[\sum_l 2<\psi|\hat L_l^{\dag }|\psi>
\hat L_l-\hat L_l^{\dag}\hat L_l
- <\psi|\hat L_l^{\dag}|\psi> <\psi|\hat L_l|\psi>\right]|\psi>,\nonumber\\
B(|\psi>)dW&=&\sum_l (\hat L_l-<\psi|L_l|\psi>)|\psi> dW_l.
\end{eqnarray}
 the QSD equation (3) assumes the standard form of a  stochastic differential equation (SDE) for an
$n$-dimensional autonomous (stationary) complex diffusion process:
\begin{equation}
d|\psi>=f(|\psi>)dt+B(|\psi>)dW.
\end{equation}
 $|\psi(t)>$ and $f(|\psi(t)>)$ are
complex vectors of complex dimension $n$, and $dW$ are
differential increments of an m-dimensional complex Wiener
process:
\begin{eqnarray}
{\rm E}[dW_l]={\rm E}[dW_ldW_{l'}]&=&0,\nonumber\\
{\rm E}[dW_l{ d\bar W}_{l'}]&=& \delta_{l,l'}dt,\nonumber\\
l&=&1,2\dots m,
\end{eqnarray}
where $E[\cdot]$ denotes the expectation with respect to the
probability distribution given by the (m-dimensional)  process
$W$,  and  $\bar W_l$ is the complex conjugate of $ W_l$.
$B(|\psi>)$ is $n\times m$ matrix, where $m$ is at most $n^2-1$, and the
diffusion matrix is
\begin{equation}
G=BB^{\dag}.
\end{equation}
Thus, $G(|\psi>)$ is Hermitian and nonnegative-definite. Notice
that, unlike the case of a general SDE, the dissipative part of
the drift (4) and  the diffusion term (5) are determined by the
same operators $\hat L_l$, and related in such a way that the
diffusion equation preserves the norm of the state vector.

The complex n-dimensional equation (3) generates
2n-dimensional real diffusion. Let us introduce the following real
$n$ dimensional vectors
\begin{eqnarray}
p&=&{i\over \sqrt{2}}(\bar\psi-\psi),\qquad q={1\over
\sqrt{2}}(\bar\psi+\psi) \nonumber\\
\psi&=&{1\over \sqrt{2}}(q+ip),\qquad \bar \psi={1\over
\sqrt{2}}(q-ip),
\end{eqnarray}
and a $2n$ dimensional vector $X=(q,p)$. Similarly, we introduce
real and imaginary parts of the vector $f$ and order them as
components of a 2n real vector ${\cal F}=(f^R, f^I)$, and
introduce real and imaginary parts of the increments of the complex
m-dim Wiener process $dW$ by
\begin{equation}
dW_i=(dW_i^R+idW_i^I)/{\sqrt 2},\> i=1,2,\dots m
\end{equation}

It is easily checked that the real and the imaginary parts are
increments of a real 2m-dimensional process, i.e.
 \begin {equation}
E(dW_i^RdW_j^R)=E(dW_i^IdW_j^I)=\delta_{i,j}dt,\qquad
E(dW_i^RdW_j^I)=0.
\end{equation}

With this notation we have
\begin{equation}
\pmatrix{dq\cr dp}={1\over \sqrt 2}\pmatrix{d\bar\psi+d\psi\cr
d\bar\psi-d\psi}.
\end{equation}
Substitution of the complex equation (3) and its complex conjugate,
leads to the following $2n$ dimensional real SDE:
\begin{equation}
\pmatrix{dq\cr dp}=\pmatrix{f^R(p,g)\cr f^I(p,q)}+{1\over \sqrt
2}\pmatrix{B^R&-B^I\cr B^I&B^R}\pmatrix{dW^R\cr dW^I},
\end{equation}

 The matrix ${\cal B}$ of
dimension $2n\times 2m$
\begin{equation}
{\cal B}={1\over \sqrt 2}\pmatrix{B^R&-B^I\cr B^I&B^R},
\end{equation}
where
\begin{equation}
(B)_{ij}=(B^R)_{ij}+i(B^I)_{ij}
\end{equation}
gives the diffusion matrix ${\cal G}$ for the real $2n$
dimensional diffusion described by the process (13)
\begin{equation}
{\cal G}={\cal B}{\cal B}^T={1\over
2}\pmatrix{(B^R)(B^R)^T+(B^I)(B^I)^T&(B^R)(B^I)^T-(B^I)(B^R)^T\cr
(B^I)(B^R)^T-(B^R)(B^I)^T&(B^I)(B^I)^T+(B^R)(B^R)^T},
\end{equation}

We can write the matrix ${\cal G}$ in terms of real and imaginary
components of the $n\times n$ complex matrix $G=BB^{\dag}$ as
follows
\begin{equation}
{\cal G}={1\over 2}\pmatrix{G^R&G^I\cr -G^I&G^R}={1\over
2}\left[\pmatrix{G^R&0\cr 0&G^R}+\pmatrix{0&-1\cr
1&0}\pmatrix{G^I&0\cr 0&G^I}\right]
\end{equation}
where  $-G^I=(G^I)^T$, since the matrix $ G$ is Hermitian.
Furthermore, one can see that, besides the equalities between the
entries corresponding to the symmetry of the matrix,
 there are other equalities
\begin{equation}
({\cal G})_{i,j}=({\cal G})_{i+n,j+n},\quad i,j=1,2\dots n
\end{equation}

 The matrix
${\cal G}$ is symmetric  and nonnegative, but it could be singular.
However, the matrix ${\rm Diag}\{1/2,1/2,\dots,1/2\}+{\cal G}$ gives a
 Riemannian metric on the real $2n$ dimensional vector space.
 The factor $1/2$ of the Euclidian part is chosen in order that the Euclidian norm of a vector
 corresponding to a complex n-vector of unit norm is also unity.

Once the diffusion metric ${\rm Diag}\{1/2,1/2,\dots,1/2\}+{\cal
G}$ is calculated the standard formulas \cite{Kobajasi} give the
connection coefficients $\Gamma^k_{\mu \nu}$ of the Levi-Civita
connection for this metric
 in terms of the coefficients $g_{\mu \nu}=\delta_{\mu \nu}/2+({\cal
G})_{\mu \nu}$ only
 \begin{equation}
\Gamma^k_{\mu \nu}={1\over 2}g^{k
\lambda}(\partial_{\mu}g_{\lambda\nu}+\partial_{\nu}g_{\lambda\mu}-\partial_{\lambda}g_{\mu\nu})
\end{equation}
 Curvature tensor, Ricci tensor and the scalar curvature of the diffusion metric
are also given by the standard formulas \cite{Kobajasi}:
\begin{eqnarray}
R^k_{\lambda\mu\nu}&=&\partial_{\mu}\Gamma^k_{\nu\lambda}-\partial_{\nu}\Gamma^k_{\mu\lambda}
+\Gamma^{\eta}_{\nu\lambda}\Gamma^{k}_{\mu\eta}-\Gamma^{\eta}_{\mu\lambda}\Gamma^{k}_{\nu\eta},\\
Ric_{\mu\nu}&=&R^{\lambda}_{\mu\lambda\nu},\quad\qquad {\cal
R}=g^{\mu\nu}Ric_{\mu\nu}.
\end{eqnarray}

Before we present the results of calculations of the diffusion
metric and its curvature for different types of environments,
 we would like to consider briefly real representation of the QSD equation in the case when the Linblad operators are Hermitian.
  This includes, for example, the dephasing environment or measurement, or the primary QSD \cite{IanProc}, \cite {QSD} and other
   fundamental stochastic modifications of the Schroedinger equation \cite{Pearle},\cite{Girardi},\cite{Hugh}.
   The goal of this digression is to point out to the connection between
   the general QSD equation (3) and  some other stochastic modifications of the Schroedinger
   equation that have the form of a norm-preserving diffusion
   equation, and that consequently the construction of the diffusion metrics and its
   properties are applicable to these equations also.
In the case of  Hermitian Linblad operators  the real
representation of (3) assumes a specially simple and illuminating
form.  Applying the same derivation as from equation (9) to
equation (13) one obtains the following:
\begin{eqnarray}
dp_i&=& -H_{ij} q_j dt+(2<L>L_{ij}-(L^2)_{ij}-<L^2>\delta_{ij})p_j dt\nonumber\\
&+&{1\over \sqrt 2}(L_{ij}-<L>\delta_{ij})p_j dW^R+{1\over \sqrt
2}(L_{ij}-<L>\delta_{ij})q_j dW^I,
\end{eqnarray}
where we have, for reasons of simplicity, included only one
Linblad operator and the summation over repeated indexes is
assumed. Noticing that for an arbitrary linear operator $B$
\begin{equation}
B_{ij}q_j=\delta_{ij} {\partial <B>\over\partial q_i}, \quad
B_{ij}p_i=\delta_{ij} {\partial <B>\over\partial p_i}
\end{equation}
equation (23) becomes
\begin{equation}
dp_i=-{\partial <H>\over \partial q_i}dt +{\partial \Delta^2
L\over \partial p_i}dt+ {1\over \sqrt 2}\left [{\partial <L>\over
\partial q_i }dW^R+{\partial <L>\over \partial p_i} dW^I\right ],
\end{equation}
where $\Delta^2 L=<L^2>-<L>^2$.
 There is an analogous equation for $dq_i$.
 The two sets of equations represent a diffusion process on $R^{2n}$, consisting
 of the drift given by a Hamiltonian dynamical
 system on $R^{2n}$ with the Hamilton's function $<H>$ and the dissipative
 part determined by $\Delta^2 L=<L^2>-<L>^2$  and the diffusion term determined by $<L>$.
  The drift and the diffusion
 are such that the norm of the vectors in $R^{2n}$ is preserved.
 Furthermore, the equations are invariant under a global gauge
 transformation corresponding to the multiplication of vectors $
 |\psi>$ by a phase factor. Takeing into the account the norm invariance and the global phase symmetry
 the equations can be written as a diffusion equation on the phase space
 $S^{2n-1}/S^1$ of the following form
\begin{equation}
dX=\Omega \nabla <H>dt+\nabla (\Delta^2 L)dt+{1\over \sqrt 2}
\nabla <L>dW
\end{equation}
where $\nabla$ and $ \Omega \nabla$ are the gradient and the skew
gradient on $S^{2n-1}/S^1$, and $X$ denotes the set of $2n-2$
coordinates on the reduced phase space $S^{2n-1}/S^1$.
 Equations like (25) have been analyzed as candidates for a description of the spontaneous state
  reduction in \cite{Adler}, or in the case $\hat L=\hat H$ in
  \cite{Hugh}.

\section{ QSD metric and qualitative properties of dynamics}

Application of  formula (16) gives for the case (5)  of the QSD
equation an explicit procedure
 for calculation of the diffusion metric coefficients $g_{ij}$, in terms of the coefficients of the Linblad
 operators and the coefficients of the state $\psi>$ in some bases $|\psi>=\sum_i c_i|i>$.
 The components of the diffusion matrix $G=BB^{\dagger}$ are given by
\begin{equation}
B_{kk'}(c,\bar c)=\sum_l^m \sum_{j,j'}^n (L^l-<L^l>{\bf 1})_{kj}
(L^{l \dagger}- <L^{l\dagger}>{\bf 1})_{k'j'} c_j\bar c_j'
\end{equation}
where: $<L^l>=\sum_{s s'}L_{s s'}c_s'\bar c_s$. Expressing
$c_i,\bar c_i$ in terms of $x_1\dots x_{2n}$
\begin{eqnarray}
x_i=(\bar c_i+ c_i)/\sqrt {2}\quad i=1,\dots n\nonumber\\
x_i=\sqrt{-1}(\bar c_i- c_i)/\sqrt {2}\quad i=n+1,\dots 2n,
\end{eqnarray}
 separating of $G^R$ and $G^I$ and substituting in (16) finally gives the $4n^2$  entries of the real matrix ${\cal G}$.

 We shall study the diffusion metric for the following three
types of environments: (a) dephasing environment; (b) the
environment corresponding to measurement of an observable and (c)
 thermal environment. The first two
 are represented by Hermitian and the third one by a non-Hermitian Linblad operators.
  The main geometrical object which we shall study are the diffusion metric norm of a state
   vector and its scalar curvature. In order to illustrate how these objects depend on the environment
    we shall use the simplest but important quantum system, namely a single qubit.
    The system operators can be expressed as combinations of the Pauli
    sigma matrices $\hat \sigma_x,\hat\sigma_y,\hat\sigma_z$, a state $|\psi>$ of unit norm is determined by
$<\hat \sigma_x>,<\hat \sigma_y>,<\hat \sigma_z>$ or by the spherical angles $(\theta,\phi)$
 given by
\begin{eqnarray}
<\hat \sigma_z>&=&\cos ( \theta)\nonumber\\
<\hat \sigma_x>&=&\sin ( \theta)\cos(\phi)\nonumber\\
<\hat \sigma_y>&=&\sin (\theta)\sin( \phi),
\end{eqnarray}
  The environment operators are \cite{env},\cite{Carmicele}
\begin{equation}
\hat L=\mu \hat\sigma_+\hat\sigma_-
\end{equation}
for the dephasing and
\begin{equation}
\hat L=\mu_1 \hat\sigma_+ +\mu_2\hat\sigma_-
\end{equation}
for the thermal environment, with $\mu_1$ and $\mu_2$  proportional to the temperature,
and finally for the measurement of, say,  $\hat\sigma_z$ the Linblad operator is just
\begin{equation}
\hat L=\mu\hat\sigma_z  .
\end{equation}

The formulas for the entries $g_{ij}$ of the diffusion metrics in terms of the
 coordinates $x_1,x_2,x_3,x_4$  in the three considered cases can be conveniently written using the following notation:
\begin{eqnarray}
d_1^2&=&x_1^2+x_3^2,\quad d_2^2=x_2^2+x_4^2 \nonumber\\
s&=&x_1x_2+x_3x_4, \quad a=x_1x_4-x_2x_3.
\end{eqnarray}
Because many of the metric entries are repeated, it is more convenient
to present them in a list rather than to write down the corresponding matrices.
 Using the notation (32), the entries of the metrics in the three
 considered cases are: For dephasing:
\begin{eqnarray}
g_{11}&=&1/2+(\mu^2/16)d_1^2(2+d_1^2)^2,\> g_{12}=(\mu^2/16)sd_1^2(2+d_1^2),\>
g_{13}=0\nonumber\\
g_{14}&=&(\mu^2/16)ad_1^2(2+d_1^2),\>
g_{22}=1/2+(\mu^2/16)d_1^4d_2^2,\>
g_{23}=-g_{14}\nonumber\\
 g_{24}&=&0,\> g_{33}=g_{11},\>
g_{34}=g_{21},\> g_{44}=g_{22}; \end{eqnarray}
For the thermal environment:
\begin{eqnarray}
g_{11}&=&1/2+d_2^2[d_1^2\mu_1^2+(2+d_1^2)^2\mu_2^2]/16,\nonumber\\
g_{12}&=&s[d_1^2(2+d_2^2)\mu_1+(2+d_1^2)d_2^2\mu_2]/16,g_{1,3}=0\nonumber\\
g_{14}&=&a[d_1^2(2+d_2^2)\mu_1+(2+d_1^2)d_2^2\mu_2]/16,\nonumber\\
g_{22}&=&1/2+d_1^2[d_2^2\mu_2^2+(2+d_2^2)^2\mu_1^2]/16\nonumber\\
g_{13}&=&0,\>g_{23}=-g_{14},\> g_{24}=0,\> g_{33}=g_{11},\>
g_{34}=g_{21},\> g_{44}=g_{22}; \end{eqnarray}
and for the measurement of $\hat \sigma_z$
\begin{eqnarray}
g_{11}&=&1/2+(g^2/16)d1^2(2+d_1^2-d_2^2)^2,\nonumber\\
g_{12}&=&(g^2/16)s(d_2^2-d_1^2-2)(d_1^2-d_2^2+2),\>
g_{13}=0\nonumber\\
g_{14}&=&a(d_1^2+d_2^2+d_1^2d_2^2),\>
g_{22}=1/2+(g^2/16)d_2^2(2+d_2^2-d_1^2)^2,\nonumber\\
g_{23}&=&-g_{14},\> g_{24}=0,\> g_{33}=g_{11},\>
g_{34}=g_{21},\> g_{44}=g_{22}.
 \end{eqnarray}

These formulas are used to compute the diffusion metric norm and the scalar curvature as functions of
the state parameters $\theta$ and $\phi$. We shall first consider the dependence of the stated properties
 of the diffusion metric on the type of the environment and the coupling strengths $\mu,\mu_1,\mu_2$ and
 then analyze the relation between these properties and the stochastic dynamics of the state vectors.

\begin{figure}
  \includegraphics[height=.5 \textheight, width=0.5 \textheight]{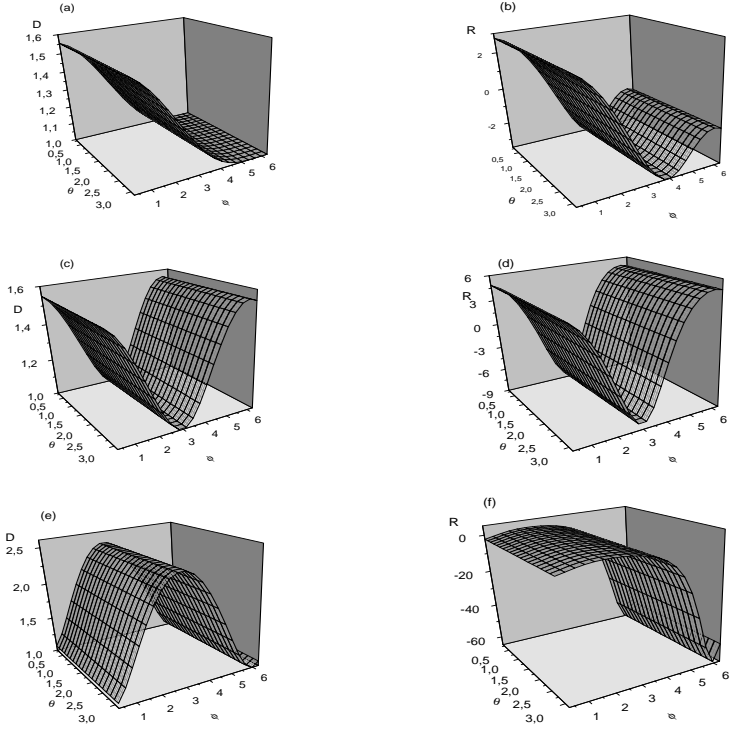}
  \caption{ Poincar\' e sections for the separability
 constrained non-symmetric quantum dynamics (39). The parameters are
$\omega=1,h=1.5$ and (a) $\mu=1.3$, (b) $\mu=1.7$}
\end{figure}

In Figures 1 and 2 we illustrate the diffusion metric norm and
curvature considered as  functions  on the sphere of states fixed
by the type of environment and the value of the corresponding
coupling $\mu,\mu_1,\mu_2$. Consider first Figure 1. The first row
(fig. (a),(c),(e))  represent the diffusion metric norm and the
second row   (fig. (b),(d),(f)) the curvature for the three types
of the environments and for
 some typical fixed values of the corresponding coupling strengths.
 The curvature is not constant, and can be positive or negative depending on the state vector and on the coupling strength.
The maxima of the curvature can be sharp like in the cases
 of the dephasing and measurement of $\sigma_z$.
 On the other hand, in the thermal case the maxima is surrounded by a
 large neighborhood of states with almost maximal value of the curvature.
Thus, the curvature has a
 sharp maxima only at the states which are clearly favored by the environment.
If there are no such states the curvature maximum differs very little from the neighboring values.
The curvature minima are at the states that are like repellers for the environment dominated dynamics.

\begin{figure}
  \includegraphics[height=.4 \textheight, width=0.5 \textheight]{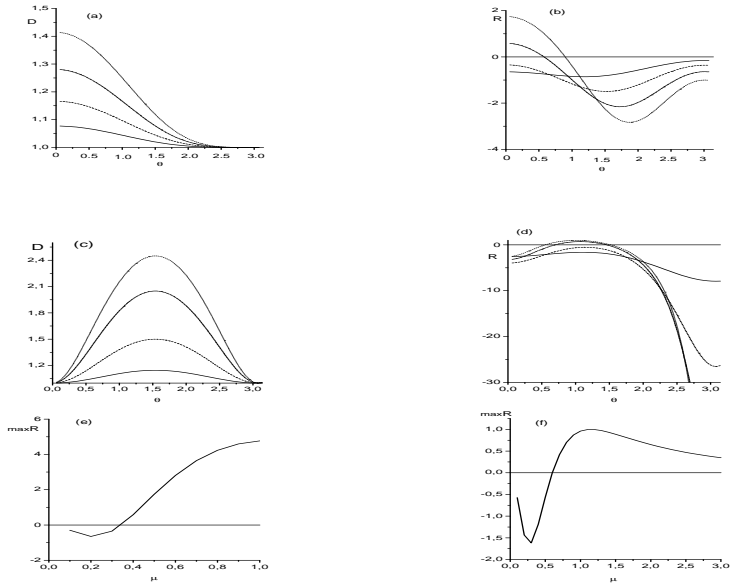}
  \caption{ Poincar\' e sections for the separability
 constrained non-symmetric quantum dynamics (39). The parameters are
$\omega=1,h=1.5$ and (a) $\mu=1.3$, (b) $\mu=1.7$}
\end{figure}

Dependence of the curvature maxima and the  norm on the coupling
strength is illustrated in Figure 2
 for the dephasing and the thermal environments.
 The most important information from these Figures is that in the dephasing and measurement ( not shown)
   cases there are clearly  sharp values of corresponding coupling strength where the curvature maxima goes
   from negative to positive values.  Also, we see that the curvature minima are negative for all values of the coupling strength.

We shall now study the relation between the sign of the curvature maxima and a stability of the stochastic
 dynamics  of the state vector. The relation will not be analyzed in a mathematically rigorous way using an
  appropriate notion of the stochastic stability and considering the  evolution of the metric as a stochastic
   process governed by the process $|\psi(t)>$. Instead,
 our strategy is to compute the curvature along different sample paths
and see if the path remains near the state corresponding to the
curvature maxima.
  We do such computations for the evolution governed by the environment
 and an additional fixed small hamiltonian, and we pay special
 attention to the case when the Linblad operators and the
 hamiltonian do not commute.
  The computations are repeated  for the values of the coupling to the environment slightly above and below
  the critical value when the curvature maxima is zero.
 If the Hamiltonian perturbation is zero the sample paths that
 started near a maximum of the curvature
  remain near this maximum.  For very small added Hamiltonian part and for a fixed value of
 the coupling to the environment,
  the  sample paths of the system could wonder away from the maximum  or could remain  near it. In the former
   case we shall say that the stochastic dynamics is unstable and in the later case
 it is stable.
   The relevant computations are illustrated in Figures 3 and 4.

\begin{figure}
  \includegraphics[height=.3 \textheight, width=0.6 \textheight]{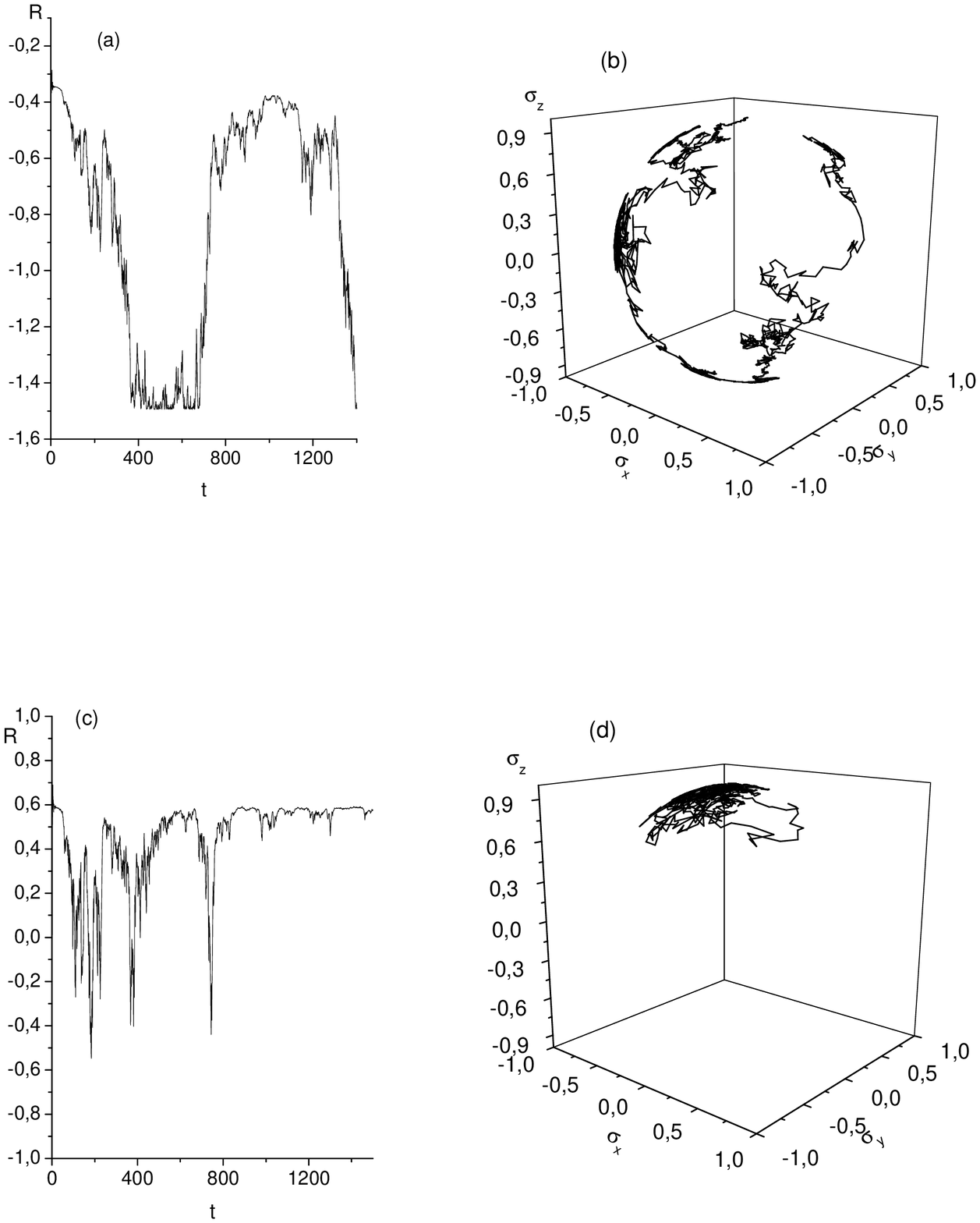}
  \caption{ Poincar\' e sections for the separability
 constrained non-symmetric quantum dynamics (39). The parameters are
$\omega=1,h=1.5$ and (a) $\mu=1.3$, (b) $\mu=1.7$}
\end{figure}

In the case of the dephasing environment (or the measurement of
$\hat\sigma_z$), when the maxima of the curvature are sharply
picked,  Figures 3  clearly illustrate that positive curvature
maxima correspond
 to the stability and negative to instability in the above mentioned sense. On the other hand, in the thermal case,
  the dynamics is always unstable even if there is no Hamiltonian perturbation. This is illustrated in Figure 4.
   We can conclude that the diffusion metric curvature provides us with a clear picture of the  qualitative
   properties of the system's dynamics under  strong influence of the environment.

\begin{figure}
  \includegraphics[height=.3 \textheight, width=0.6 \textheight]{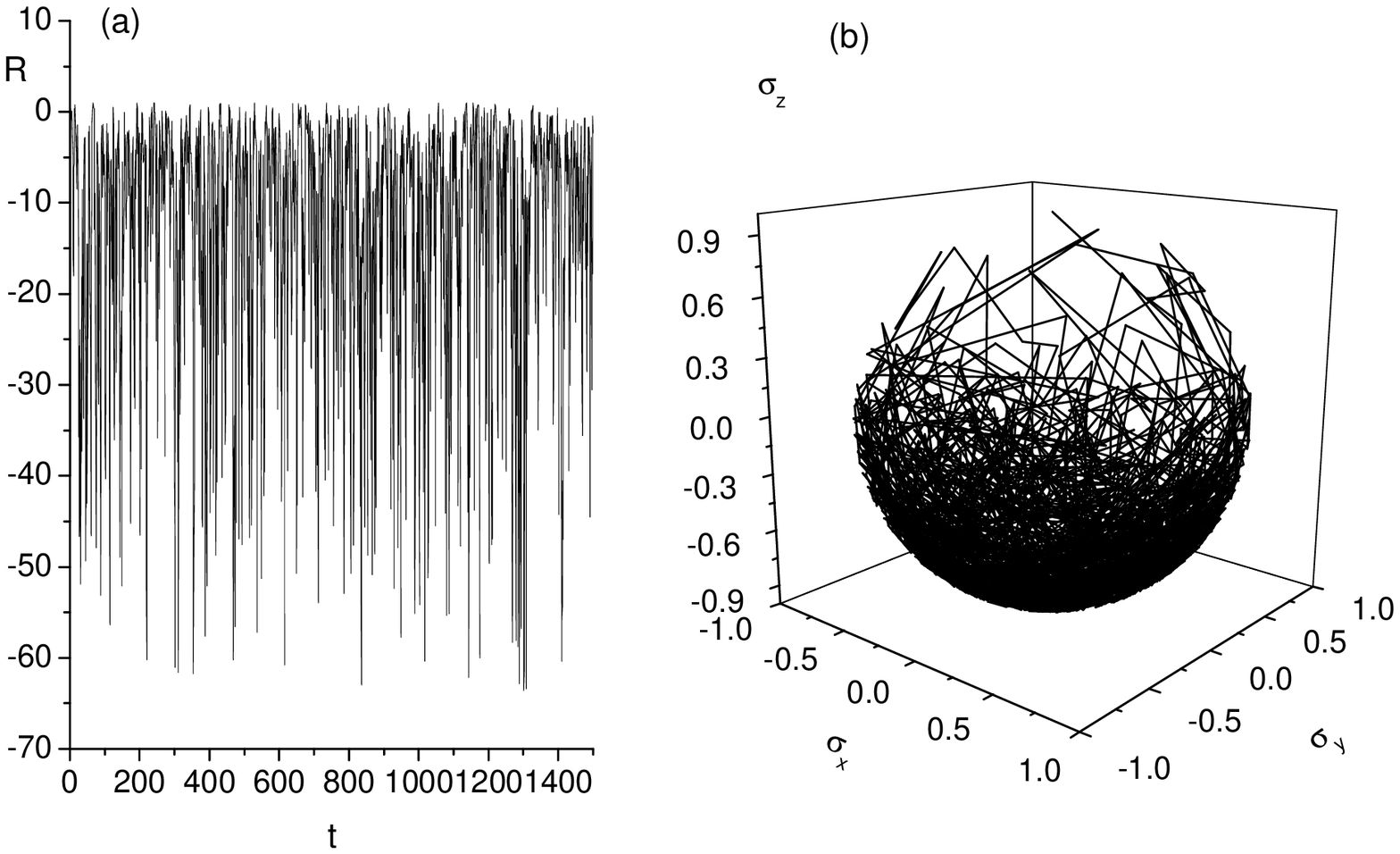}
  \caption{ Poincar\' e sections for the separability
 constrained non-symmetric quantum dynamics (39). The parameters are
$\omega=1,h=1.5$ and (a) $\mu=1.3$, (b) $\mu=1.7$}
\end{figure}

It is well known that
 if the Linblad operators are Hermitian and commute
 with the hamiltonian, than the attractors of the stochastic QSD dynamics are
 the common eigenstates of the Hamiltonian and the Linblad operators \cite{QSD},\cite{Adler}. The
 curvature maxima coincide with the eigenstates of the Linblad
 operators, and consequently with the eigenstates of the
 hamiltonian.  The
 probability of convergence to one of the attractors is, in this case, determined
 solely by the distance of the initial state from the attractor
 eigenstate, that is by the quantum mechanical transition
 probability, and does not depend on the parameters of the Hamiltonian and stochastic terms.
  The sign of the curvature maxima has no effect on this probability.
 This is the reason why we expected that the relevance of the sign of the curvature
 maxima on the stochastic stability is manifested if
 the Linblad operator and the Hamiltonian perturbation do not commute.
   This
 expectation is qualitatively confirmed, as we described and illustrated in
 Figures 3 and 4, by numerical computations.
 Observations of numerical sample paths,
  when the Linblad and the Hamiltonian operators do not commute, are enough
 to establish the qualitative connection between the maxima of the
 curvature and the stability of small domains near the maxima.

Finally, our treatment of the relation between the geometry of the
diffusion and the stability of the stochastic dynamics is rather
heuristic. We treated the diffusion metric as a given field on
$R^{2n}$ (determined by the Linblad operators), and we numerically
studied the paths of the stochastic process $|\psi(t)>$ in
relation to the sign of the curvature maxima. However, the problem
of stability versus the properties of the diffusion metric should
be formulated and studied using the appropriate notions of
stochastic
 stability \cite{Kasm},\cite{Arnold}. Nevertheless, we think that
 the numerical evidence strongly indicates that there is a clear
 relation between the sign and the shape of the curvature maxima and
 the systems dynamical stability.

\section{Summary and discussion}

According to the view of QSD theory, evolution of a state of an
open quantum system is a diffusion process governed by a complex
stochastic differential equation on the Hilbert space of the
system. The diffusion term of the QSD evolution equation
explicitly depends on the operators modelling the environment and
on the current state vector of the system. We have studied the
Riemannian metric associated with the diffusion term in the QSD
equation. The metric is defined on the real 2n-dimensional space
(here n is the complex dimension of the Hilbert space) and is
directly related to the properties of the Linblad operators of the
environment. We have shown that the scalar curvature of the metric
has local maxima at states that are favored by the corresponding environment.
 The curvature at
different points, and in particular its local maxima, can be be
negative or positive depending on the strength of the coupling to
the environment. Also, the sharpness of the curvature maxima
reflects the type of the environment. We have shown that there is
a sense in which the sign of the curvature maxima is related to
the stability of the corresponding state under the addition of a
small perturbation that does not commute with the considered
Linblad operator. If the environment type and the coupling
strength are such that the curvature has sharp positive maxima,
than the corresponding state is likely to attract the states of
the system whose evolution is governed by the environment and a
Hamiltonian that do not necessarily commute. On the other hand, if
the curvature maxima are negative, the  corresponding states are
dynamically unstable under a small  Hamiltonian perturbation that does not commute with the Linblad operators.
In conclusion, the curvature of the diffusion metric is a relatively  easy to calculate, and a
 very good indicator of what the environment dominated dynamics of the system would look like.

The QSD equation describes the evolution of a pure quantum state
using the Hilbert space of the quantum system, but, because it is
norm-preserving, it gives also an equation on the state space,
namely on the space of rays of the Hilbert space. The Riemannian
metric associated with the diffusion on $R^{2n}$ gives a Hermitian
modification of the Fubini-Study metric on $CP^{n-1}$. It is
common to consider the complex projective manifold with the
associated Fubini-Study metric as the proper framework for the
geometry of quantum states \cite{Hugh2},\cite{book}, so the
modification of the metric due to the diffusion should also be
formulated within this framework.

The examples that we have analyzed in this paper are restricted on
a single qubit under the influence of various types of
environments. It would be interesting to analyze the properties of
the diffusion metric in the case of coupled gubits, and in
particular to see what is the curvature at the entangled states.
Probably the proper framework for such analyzes is the formulation
on $CP^{n-1}$,  mentioned in the previous paragraph, because the
entangled states then have characteristic geometric interpretation
\cite{Hugh2}.

 \vskip 1cm

 {\bf Acknowledgements} This work is
partly supported by the Serbian Ministry of Science contract No.
141003. I should also like to acknowledge the support and
hospitality of the Abdus Salam ICTP.

\newpage
{\bf FIGURE CAPTIONS}

{\bf  Figure 1} Diffusion metric norm (a,c,e) and curvature (b,d,f) as functions of state parametrized by $(\theta,\phi)$,
  for dephasing environment
 with $\mu=0.6$ (a,b); measurement of $\hat\sigma_z$ with $\mu=1.$ (c,d) and thermal environment
 with $\mu_1=2,\mu_2=1$ (e,f).

{\bf  Figure 2} Diffusion metric norm (a,c) and curvature (b,d) as functions of $\theta$ for different values
 of the parameters $\mu$ or $\mu_1,\mu_2$, and the maximum over $(\theta,\phi)$ of the curvature as a function of  $\mu$ (e)
   or $\mu_1=2\mu_2=2\mu (f)$ . Figures a,b,e corespond to the dephasing and $c,d,f$ to the thermal environment.

{\bf  Figure 3}   Diffusion metric curvature (a,c) along the corresponding stochastic path illustrated in b,d
 for the dephasing environment and $\mu=0.3$ when ${\rm max} R<0$ (a,b),  and $\mu=0.5$ when ${\rm max} R>0$(c,d).
  The small Hamiltonian perturbation is
 $0.01\hat\sigma_x$.

{\bf  Figure 4}   Diffusion metric curvature (a) along a stochastic path (b) for the
thermal environment and $\mu_1=2\mu_2=1.6$. The Hamiltonian part is zero.
\end{document}